\documentclass[aps,prl,reprint,superscriptaddress,floatfix]{revtex4-2}

\usepackage{amsmath,amssymb}
\usepackage{bm}
\usepackage{graphicx}

\begin{document}

\title{Phonon-Programmable Hidden Unconventional Magnetism in Two-Dimensional Spin-Degenerate Antiferromagnets}

\author{Xiaonong Shen}
\affiliation{Physics Department, Materials Genome Institute, State Key Laboratory of Advanced Refractories, Institute for Quantum Science and Technology, Shanghai University, Shanghai 200444, China}

\author{Cheng Tang}
\affiliation{Physics Department, Materials Genome Institute, State Key Laboratory of Advanced Refractories, Institute for Quantum Science and Technology, Shanghai University, Shanghai 200444, China}

\author{Wei Ren}
\email{renwei@shu.edu.cn}
\affiliation{Physics Department, Materials Genome Institute, State Key Laboratory of Advanced Refractories, Institute for Quantum Science and Technology, Shanghai University, Shanghai 200444, China}

\date{\today}

\begin{abstract}
Spin-degenerate antiferromagnets can host hidden unconventional magnetism in
their lattice degrees of freedom. We show that coherent phonons activate this
magnetism by removing the spin-layer operations that enforce equilibrium band
degeneracy without changing the collinear N\'eel order. The frequency and
polarization of a pump electric field select a resonant \(\Gamma\)-point optical
phonon and, within a doublet, its coordinate direction. This choice fixes the
residual spin-layer symmetry. In
monolayer MnPSe$_3$, an \(A_{2u}\) mode produces \(i\)-wave splitting odd in the
mass-weighted phonon coordinate $Q$, reversing sign under $Q\to-Q$,
whereas two orthogonal directions of the same doubly degenerate \(E_u\) doublet
produce \(d\)- and \(s\)-wave splitting at a common resonance. Rotating the
in-plane pump field \(\bm E_\parallel\) therefore programs both the
  spin-splitting texture and the thermoelectric spin current, continuously tuning the
  response between transverse pure-spin and longitudinal spin-polarized
  currents. A complete classification of two-dimensional collinear spin layer
  groups identifies the \(\Gamma\)-point coordinates that remove the
  degeneracy-enforcing operations and activate such
unconventional magnetism.
\end{abstract}

\maketitle

\textit{Introduction.} Unconventional collinear magnets combine compensated
magnetic order with nonrelativistic spin-split bands~\cite{Smejkal2022Beyond,Liu2022SpinGroup,Yuan2024Gamma}. When no spin-space-group
operation enforces degeneracy between opposite-spin sublattices,
momentum-dependent spin splitting can emerge without a net magnetization or
spin-orbit coupling~\cite{Liu2022SpinGroup}. Altermagnets and compensated ferrimagnets are
two representative classes of such unconventional
magnetism~\cite{Smejkal2022Beyond,Yuan2024Gamma}. The allowed transport
response depends on their spin symmetry and may include anisotropic
spin-polarized conductivity, pure spin currents, or spin-thermoelectric
transport~\cite{Naka2019SpinCurrent,Dou2025SpinConductivity,Yi2025Nernst,Sheoran2026SpinCurrents}. By
contrast, spin-degenerate antiferromagnets retain \(\mathcal{PT}\),
spin-flipping translations, or other operations that lock the opposite-spin
bands. 

Spin-degenerate antiferromagnets can be converted into unconventional magnetic
states by external control, provided that the operations enforcing
full-Brillouin-zone spin degeneracy are broken. In two-dimensional
antiferromagnets, twisting can reconstruct the spatial symmetry of a bilayer
and generate altermagnetism~\cite{He2023TwistedBilayers}. Sliding changes the
interlayer stacking registry and thereby switches the spin-splitting wave
class~\cite{Diao2026Interwave}. Static electric fields can likewise remove the
electronic equivalence between opposite-spin sublattices and
induce altermagnetism~\cite{Mazin2023FieldMnPSe3}. Together, these studies show how
structural distortions and static fields create and control unconventional
magnetic states. A more fundamental question then arises: do the intrinsic
lattice degrees of freedom of a spin-degenerate antiferromagnet already
contain hidden unconventional magnetism? If so, how can these states be
identified and selectively driven?

Phonons provide a symmetry-resolved way to identify and activate this hidden
unconventional magnetism. Each phonon mode has a definite symmetry and
resonance frequency, allowing the pump frequency to select the resonant mode
and the pump polarization to select the corresponding phonon coordinate. A
finite mode amplitude produces a time-dependent structural distortion that
can break specific spin-degeneracy-protecting operations on picosecond time
scales~\cite{Forst2011Nonlinear}. Recent calculations have predicted that
coherent phonons transform the \(g\)-wave altermagnets MnTe and CrSb into
compensated ferrimagnetic states with global spin
splitting~\cite{Wang2026Ultrafast}. Nonlinear phononics has also been predicted
to induce Zeeman-type spin splitting in NiO~\cite{Yuan2026Dynamical}. These
results show that lattice motion can couple to
nonrelativistic spin splitting. A unified symmetry criterion is still lacking
for identifying and programming hidden unconventional
magnetism across two-dimensional spin-degenerate antiferromagnets.

In this Letter, we formulate such a symmetry criterion for activating and programming
hidden unconventional magnetism in two-dimensional
spin-degenerate antiferromagnets without changing their collinear N\'eel order.
We label textures with zero, two, four, and six
symmetry-enforced spin-degenerate nodal lines through \(\Gamma\) as \(s\)-,
\(d\)-, \(g\)-, and \(i\)-wave, respectively~\cite{Smejkal2022Beyond,Yuan2024Gamma}.
Using monolayer MnPSe$_3$ as a prototype, first-principles calculations and
symmetry analysis show that the pump frequency and field orientation select
the \(A_{2u}\) or \(E_u\) mode, while the in-plane pump polarization further
selects the \(Q_d\) or \(Q_s\) direction within the same \(E_u\) doublet. The
\(A_{2u}\) coordinate and the two \(E_u\) directions retain different
spin-group operations and produce \(Q\)-odd \(i\)-, \(d\)-, and \(s\)-wave spin
splitting, respectively. These splittings follow the coherent lattice motion
on picosecond time scales. At the common \(E_u\) resonance, rotating the pump
  polarization further tunes the spin-thermoelectric response between
  transverse pure-spin and longitudinal spin-polarized currents. A classification of all
two-dimensional collinear spin layer groups identifies the \(\Gamma\)-point
structural coordinates that can remove the degeneracy-enforcing operations and activate these unconventional magnetic
states.
\begin{figure}[t]
  \includegraphics[width=\columnwidth]{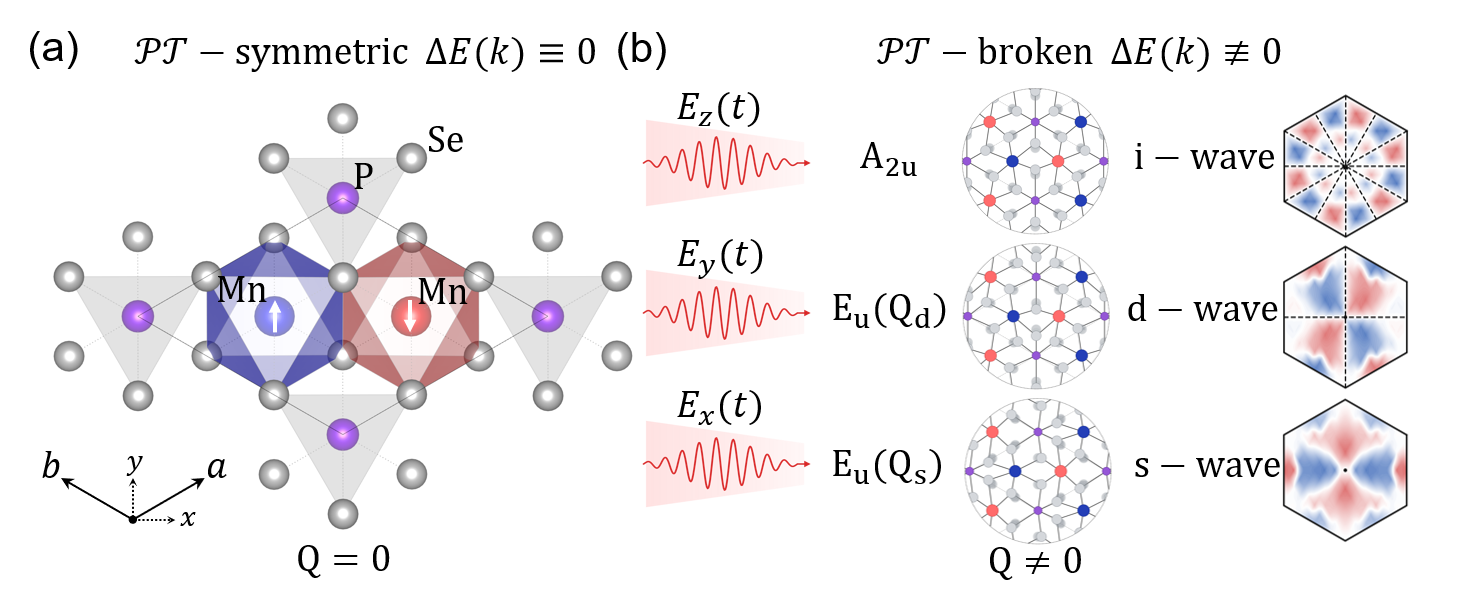}
  \caption{\label{fig:mechanism}Phonon-controlled symmetry selection in
  monolayer MnPSe$_3$. (a) Crystal structure and collinear N\'eel spin
  configuration of the undistorted, spin-degenerate state at $Q=0$. (b) Pump-polarization
  selection of $A_{2u}(Q)$, $E_u(Q_d)$, and $E_u(Q_s)$ distortions and their
  $i$-wave, $d$-wave, and $s$-wave spin-split states.}
\end{figure}

\begin{figure*}[t]
  \centering
  \includegraphics[width=0.9\textwidth]{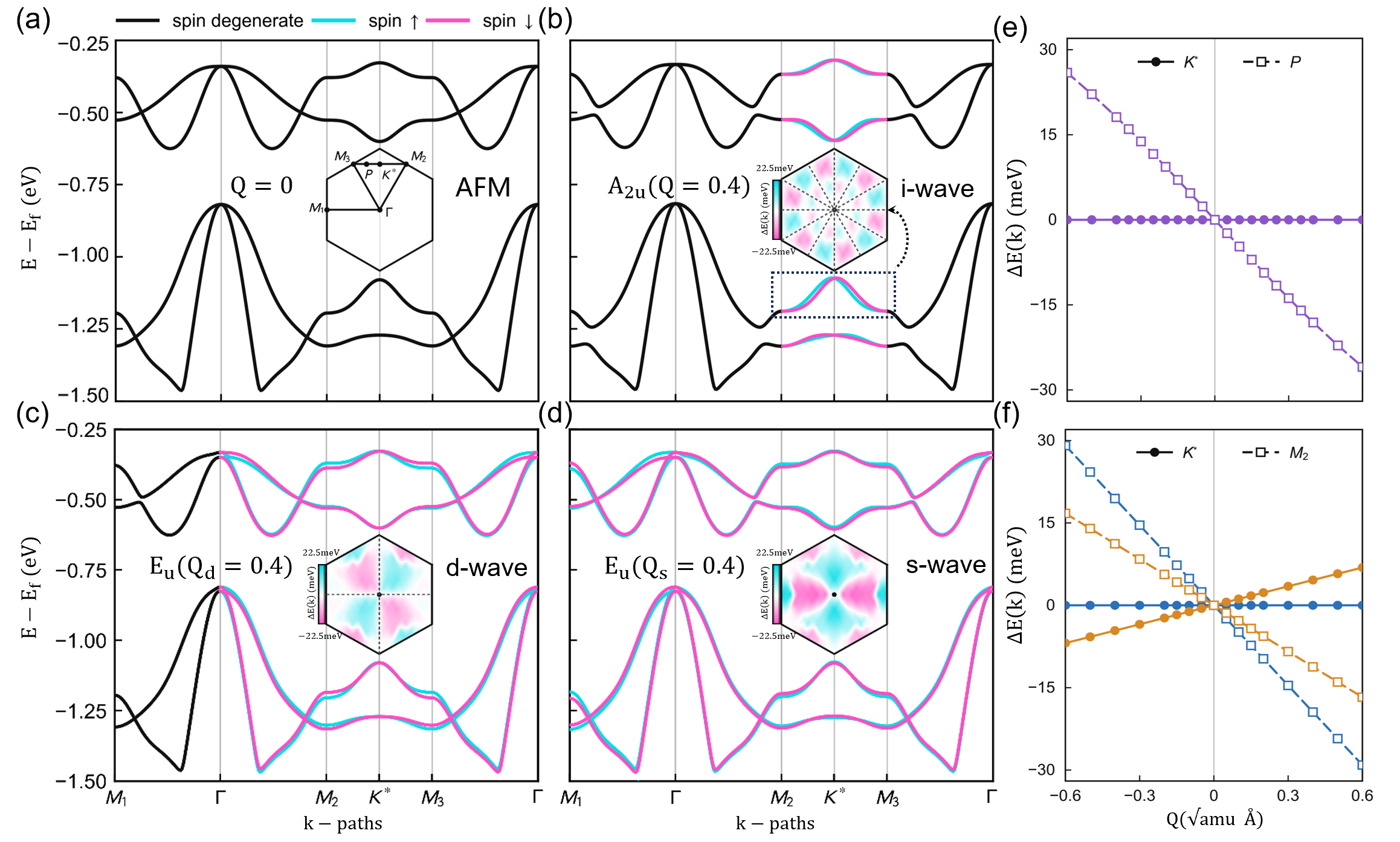}
  \caption{\label{fig:bands}Phonon-selected spin splitting in monolayer
  MnPSe$_3$. (a) Spin-degenerate bands of the undistorted $Q=0$ state.
  (b)--(d) Spin-resolved bands and Brillouin-zone splitting textures for
  $A_{2u}$, $E_u(Q_d)$, and $E_u(Q_s)$ at
  $Q=0.40\sqrt{\mathrm{amu}}\,\text{\normalfont\AA}$. Cyan and magenta denote
  the two spin channels, and black dashed lines mark symmetry-enforced nodes;
  the frame in (b) identifies the band pair used for the splitting plots.
  (e) Signed splitting at $P$ and $K^*$ versus the $A_{2u}$ coordinate.
  (f) Corresponding values at $M_2$ and $K^*$ for the two $E_u$ directions;
  blue and orange denote the $d$-wave $E_u(Q_d)$ and $s$-wave $E_u(Q_s)$
  states, respectively. Symbols are direct calculations and curves are
  $Q$-odd fits.}
\end{figure*}
\textit{Symmetry and mode selection.}---Monolayer MnPSe$_3$ crystallizes in
layer group $p\bar{3}1m$ (LG 71), with point group $D_{3d}$. The Mn$^{2+}$ ions
form a honeycomb lattice and are octahedrally coordinated by Se, while P--P
dimers occupy the honeycomb centers within $[\mathrm{P}_2\mathrm{Se}_6]^{4-}$
units [Fig.~\ref{fig:mechanism}(a)]~\cite{Sivadas2016Gate}. Its magnetic ground
state is a collinear N\'eel antiferromagnet with opposite spins on the two Mn
sublattices~\cite{Ni2021MnPSe3Neel,Mai2021MnPSe3MagnonPhonon}. Bulk MnPSe$_3$
orders below $T_N\simeq74$~K, while long-range N\'eel order persists in
monolayers with $T_N\simeq40$~K
~\cite{Mai2021MnPSe3MagnonPhonon,Ni2021MnPSe3Neel}. For a two-dimensional
magnet, the allowed momentum-space spin splitting is determined by its spin
layer group. Suppressing
lattice translations, the spin layer group of the $Q=0$ state is
\begin{equation}
\mathcal G_0=[E\parallel D_3]+[U\parallel\mathcal I D_3],
\label{eq:q0-group}
\end{equation}
where $E$ is the identity spin operation and $U=C_{2\perp}$ reverses the
collinear spins. The spatial subgroup $D_3$ preserves each Mn spin sublattice,
whereas $\mathcal I D_3=\{\mathcal I,S_{6z}^{+},S_{6z}^{-},M_1,M_2,M_3\}$
exchanges them. In the nonrelativistic collinear limit,
$[U\parallel\mathcal I]$ represents the $\mathcal{PT}$ symmetry of the $Q=0$
state. It relates $(\uparrow,\bm k)$ to
$(\downarrow,-\bm k)$, and the momentum evenness of each spin channel then
gives $E_{n\uparrow}(\bm k)=E_{n\downarrow}(\bm k)$ throughout the Brillouin
zone. The undistorted structure is therefore fully spin degenerate.
From a lattice-dynamical viewpoint, $Q=0$ denotes the undistorted structure.
Freezing a phonon at $Q\neq0$ lowers the structural symmetry and can remove the
operations that enforce spin degeneracy at $Q=0$. To select such distortions
optically, we consider
$\Gamma$-point phonons because a long-wavelength pump electric field couples
directly to infrared-active zone-center coordinates~\cite{Forst2011Nonlinear}.
The optical representation
of monolayer MnPSe$_3$ is
\begin{equation}
\Gamma_{\mathrm{opt}}=3A_{1g}\oplus2A_{2g}\oplus5E_g\oplus A_{1u}
\oplus3A_{2u}\oplus4E_u .
\end{equation}
We retain only modes that are infrared active and whose finite-amplitude
distortions remove every operation enforcing spin degeneracy throughout the
Brillouin zone. This selects the infrared-active $3A_{2u}\oplus4E_u$ modes.
At $\Gamma$, $A_{2u}(Q)$ is a nondegenerate mode described by a single coordinate $Q$,
whereas $E_u$ is a doubly degenerate phonon doublet described by
$\bm Q_{E_u}=(Q_d,Q_s)$. The $A_{2u}(Q)$ coordinate couples to the out-of-plane
field $E_z(t)$. In the axes
of Fig.~\ref{fig:mechanism}(b), $E_y(t)$ selects
$E_u(Q_d)\equiv E_u(Q_d,0)$, while the orthogonal field $E_x(t)$ selects
$E_u(Q_s)\equiv E_u(0,Q_s)$.
Because $Q_d$ and $Q_s$ are orthogonal directions of the same
symmetry-degenerate $E_u$ doublet, rotating $\bm E_\parallel$ changes the
spin-layer group of the distorted structure without changing the phonon branch
or resonance frequency.

The three $Q\neq0$ structures have spin layer groups
\begin{subequations}
\label{eq:distorted-groups}
\begin{align}
\mathcal G_{A_{2u}(Q)}&=[E\parallel C_3]+[U\parallel M_1C_3],
\label{eq:group-a2u}\\
\mathcal G_{E_u(Q_d)}&=[E\parallel C_1]+[U\parallel M_dC_1],
\label{eq:group-eu-d}\\
\mathcal G_{E_u(Q_s)}&=[E\parallel C_2].
\label{eq:group-eu-s}
\end{align}
\end{subequations}
Here $C_1=\{E\}$, $C_3=\{E,C_{3z}^{+},C_{3z}^{-}\}$,
$M_1C_3=\{M_1,M_2,M_3\}$, and $M_d$ is the mirror preserved by $Q_d$.
For $A_{2u}(Q)$, the threefold rotation and three spin-exchanging mirrors
enforce six nodal lines through $\Gamma$, producing an $i$-wave altermagnetic
state. The $E_u(Q_d)$ structure retains one spin-exchanging mirror, which together
with momentum evenness leaves two orthogonal nodal lines and a $d$-wave
altermagnetic state. The orthogonal $E_u(Q_s)$ structure retains only a
spin-preserving twofold rotation. It has no symmetry-enforced nodal line and
therefore realizes a compensated ferrimagnetic state with $s$-wave spin
splitting [Fig.~\ref{fig:mechanism}(b)].

For the calculations below, we choose a $6.314\,\mathrm{THz}$ $A_{2u}$ branch
with $|Z_\nu|=0.0564\,e/\sqrt{\mathrm{amu}}$ and a
$3.700\,\mathrm{THz}$ $E_u$ doublet with basis-independent in-plane
mode-effective-charge singular value $\sigma_Z=0.314\,e/\sqrt{\mathrm{amu}}$.
Frequency distinguishes the two resonances, while in-plane polarization
selects $Q_d$ or $Q_s$ within $E_u$. The complete mode-frequency and
effective-charge spectrum $Z(f)$ is given in Fig.~S1 of the Supplemental
Material~\cite{SupplementalMaterial}.

\textit{Phonon-selected spin-split states.}---Using first-principles
calculations, Fig.~\ref{fig:bands} tests these
symmetry predictions using spin-resolved bands. The two spin channels of the undistorted $Q=0$ structure are fully degenerate, with
$\Delta E(\bm k)=0$ at every momentum [Fig.~\ref{fig:bands}(a)]. At
$Q=0.40\sqrt{\mathrm{amu}}\,\text{\normalfont\AA}$, the $A_{2u}$,
$E_u(Q_d)$, and $E_u(Q_s)$ frozen-phonon structures display the predicted $i$-, $d$-, and
$s$-wave textures [Figs.~\ref{fig:bands}(b)--\ref{fig:bands}(d)]. The maximum
absolute splittings of the displayed bands across the Brillouin zone are
$20.59$, $19.44$, and $22.47\,\mathrm{meV}$, respectively.

The phonon coordinate controls both the magnitude and sign of the splitting
[Figs.~\ref{fig:bands}(e) and \ref{fig:bands}(f)]. Because $A_{2u}$ and $E_u$
are odd under inversion, the $[U\parallel\mathcal I]$ operation of the $Q=0$
structure takes the $+Q$ structure into the $-Q$ structure while exchanging the spin
channels. Together with the momentum evenness of each spin-resolved dispersion, the signed
splitting $\Delta E(\bm k,Q)=E_{\uparrow}(\bm k,Q)-E_{\downarrow}(\bm k,Q)$
obeys
\begin{equation}
\Delta E(\bm k,-Q)=-\Delta E(\bm k,Q).
\label{eq:qodd}
\end{equation}
All three frozen-phonon structures satisfy this relation. Near $Q=0$, their splitting is
linear in $Q$. Across the calculated interval
$|Q|\leq0.60\sqrt{\mathrm{amu}}\,\text{\normalfont\AA}$, the two $E_u$
states remain nearly linear, while $A_{2u}$ develops a weak nonlinear
correction without losing its odd parity. The instantaneous coordinate can
therefore reverse and continuously tune the spin splitting.

\textit{Coherent phonon dynamics.}---We next solve the resonantly driven
$A_{2u}$ and $E_u$ coordinates. The mass-weighted coordinate of mode $\nu$
obeys
\begin{equation}
C_\nu\ddot Q_\nu+2C_\nu\Gamma_\nu\dot Q_\nu
+\partial_{Q_\nu}V_\nu(Q_\nu)=Z_\nu E_\nu(t),
\label{eq:dynamics}
\end{equation}
where $V_\nu$ is the fitted even anharmonic potential and
$\Gamma_\nu=2\pi\gamma_\nu$ is obtained from the three-phonon linewidth at
$50\,\mathrm{K}$. Resonant Gaussian pulses with a $1\,\mathrm{ps}$ intensity
full width at half maximum are chosen to reach
$\max|Q_\nu|=0.398\sqrt{\mathrm{amu}}\,\text{\normalfont\AA}$, within the
frozen-phonon range. The resulting coherent-amplitude decay times are
$2.68\,\mathrm{ps}$ for $A_{2u}$ and $24.6\,\mathrm{ps}$ for $E_u$. The
detailed parameters are reported in the Supplemental
Material~\cite{SupplementalMaterial}.

\begin{figure}[t]
  \includegraphics[width=\columnwidth]{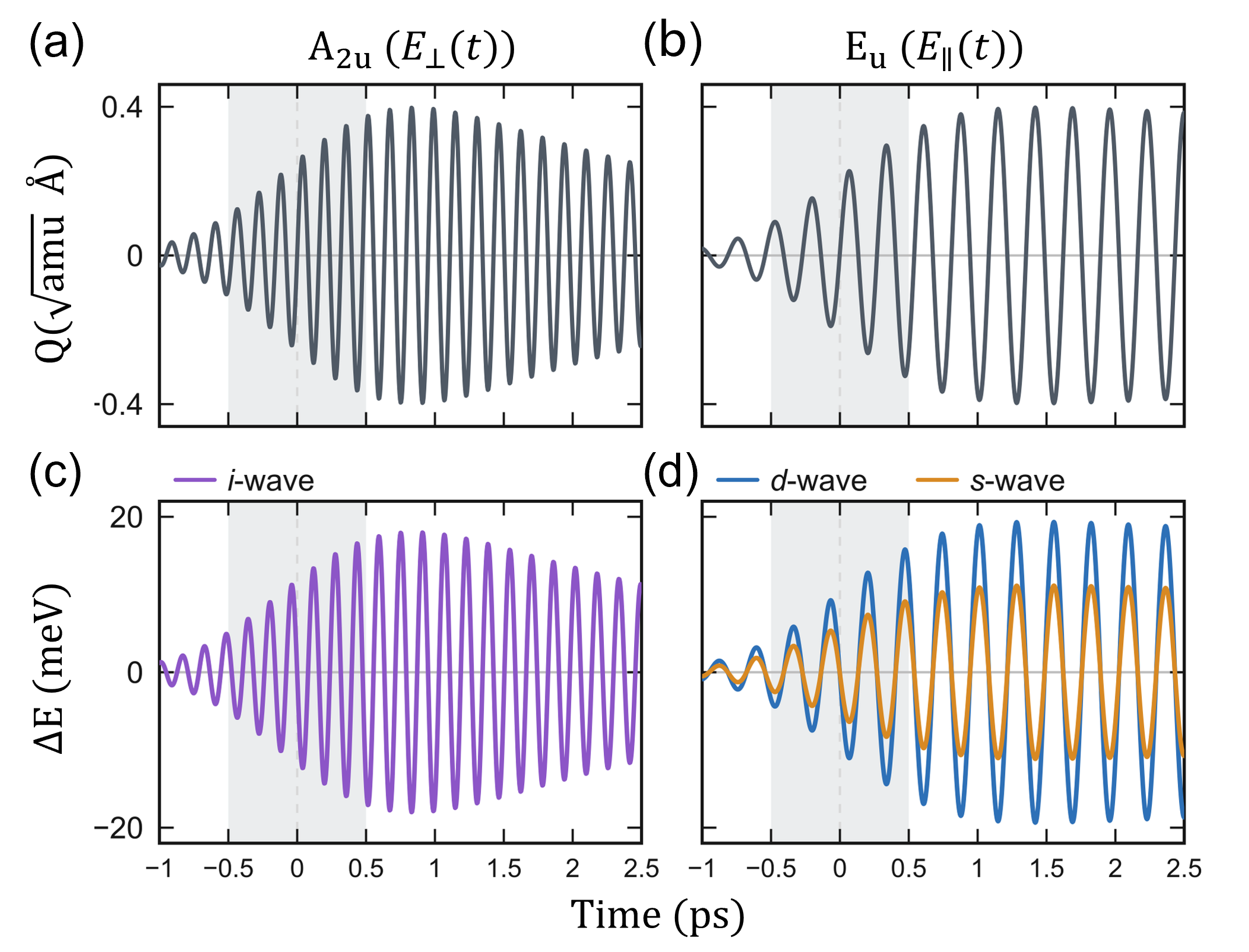}
  \caption{\label{fig:dynamics}Damped coherent-phonon dynamics and induced spin
  splitting at $50\,\mathrm{K}$. (a),(b) Driven $A_{2u}$ and $E_u$
  coordinates. The gray interval marks the pump intensity full width at half
  maximum. (c) Time-dependent $i$-wave splitting at $P$ generated by
  $A_{2u}$. (d) Time-dependent $d$- and $s$-wave splittings at $M_2$ generated
  by $E_u(Q_d)$ and $E_u(Q_s)$. Independently fitted $Q_d$ and $Q_s$
  potentials yield indistinguishable lattice trajectories on this scale, while
  their electronic responses are evaluated separately. The splitting traces use an
  adiabatic frozen-phonon interpolation, not real-time electronic propagation.}
\end{figure}

As shown in Fig.~\ref{fig:dynamics}, both coordinates build up coherently
during the pulse. The $A_{2u}$ oscillation then decays appreciably, whereas the
$E_u$ oscillation persists over the displayed interval. Applying the
frozen-phonon relation $\Delta E_\nu(t)=\Delta E_\nu[Q_\nu(t)]$ gives peak
magnitudes of $18.05\,\mathrm{meV}$ at $P$ for $A_{2u}$ and
$19.35\,\mathrm{meV}$ and $11.15\,\mathrm{meV}$ at $M_2$ for $E_u(Q_d)$ and
$E_u(Q_s)$. Full spin degeneracy is restored whenever $Q_\nu(t)$ crosses zero.
Equation~\eqref{eq:qodd} then requires the splitting to reverse in successive
half cycles.
Thus the pump selects the nodal class, while the phonon phase controls its
instantaneous sign.

\begin{figure}[t]
  \includegraphics[width=\columnwidth]{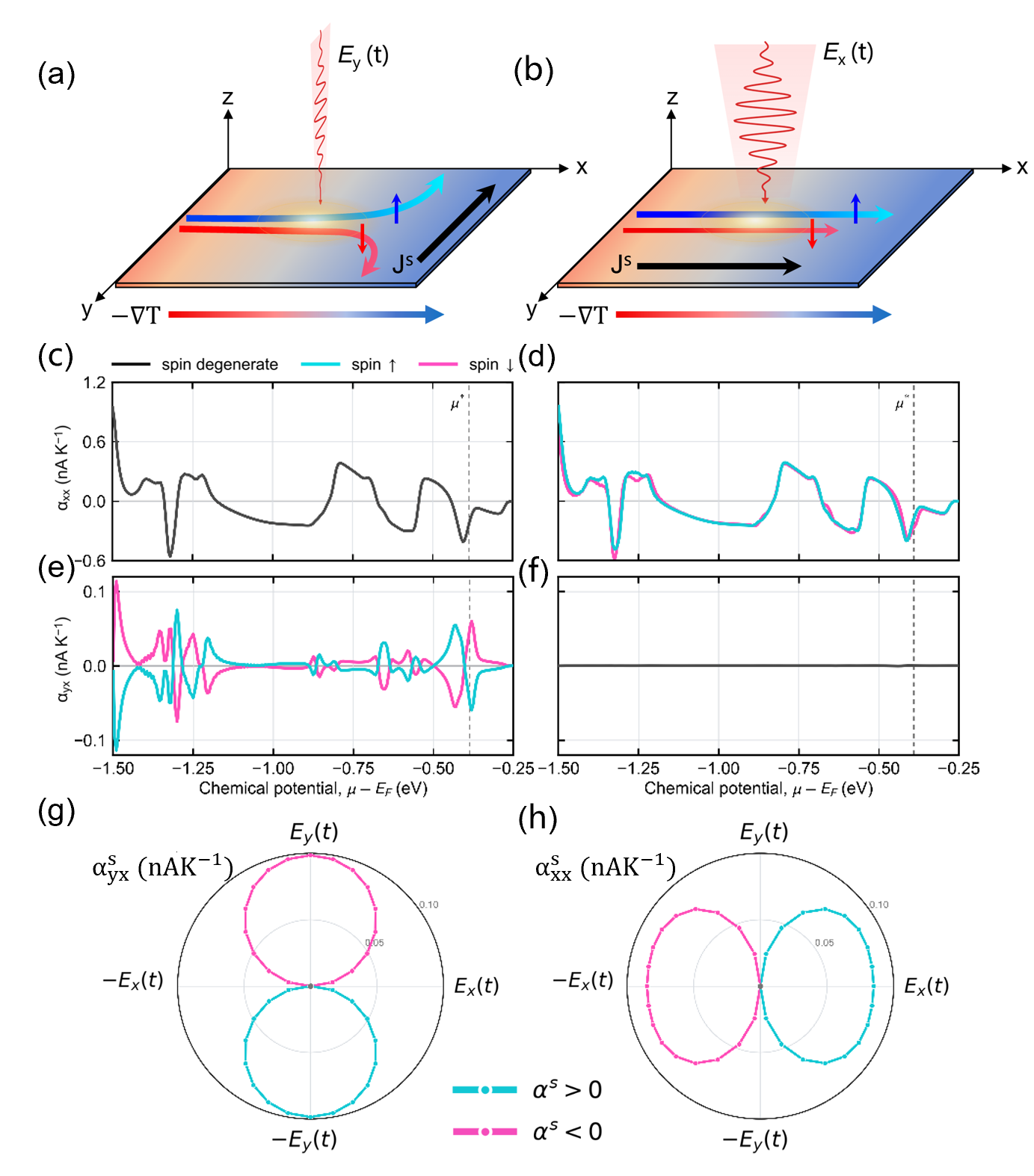}
  \caption{\label{fig:transport}Polarization-selected spin-thermoelectric
  readout. (a),(b) Transverse and longitudinal responses selected by $E_y(t)$
  and $E_x(t)$ for $-\bm\nabla T\parallel x$. (c),(d) Longitudinal
  $\alpha_{xx}^{\uparrow,\downarrow}$ and (e),(f) transverse
  $\alpha_{yx}^{\uparrow,\downarrow}$ for $E_u(Q_d)$ [(c),(e)] and
  $E_u(Q_s)$ [(d),(f)] at
  $Q=0.40\sqrt{\mathrm{amu}}\,\text{\normalfont\AA}$. Dashed lines mark
  $\mu^*=E_{\mathrm{VBM}}(Q)-0.1096\,\mathrm{eV}$. (g),(h) Angular dependence
  of $\alpha_{yx}^s$ and $\alpha_{xx}^s$ at fixed $|\bm Q|$, with polar axes
  labeled by pump-field direction.}
\end{figure}

\textit{Polarization-programmed spin-thermoelectric readout.}---In this system,
the two orthogonal directions of the $E_u$ doublet generate $d$- and $s$-wave
spin-splitting textures and give rise to spin-thermoelectric response in
different directions. The $d$-wave state produces a transverse pure spin
current, whereas the $s$-wave state produces a longitudinal spin-polarized
current. This distinction provides a transport readout of the
phonon-programmed unconventional magnetic state. By contrast, the
$A_{2u}$-induced $i$-wave state gives no in-plane spin-thermoelectric response
within the Boltzmann linear-response framework: its residual $C_3$ rotation and
three spin-exchanging mirrors force $\bm\alpha^s=0$. We therefore use the
$E_u$ mode as the transport readout. Unlike the nondegenerate $A_{2u}$ mode,
the $E_u$ doublet allows the in-plane pump polarization to select $Q_d$ or
$Q_s$ at the same resonance. We evaluate the quasistatic response with the
collinear N\'eel electronic structure held fixed. For a temperature gradient
$-\bm\nabla T$, we define
$J_i^{c,s}=\alpha_{ij}^{c,s}(-\nabla_jT)$, with
$\alpha_{ij}^{c}=\alpha_{ij}^{\uparrow}+\alpha_{ij}^{\downarrow}$ and
$\alpha_{ij}^{s}=\alpha_{ij}^{\uparrow}-\alpha_{ij}^{\downarrow}$.
The $E_u(Q_d)$ structure retains the spin-exchanging mirror
$[U\parallel M_d]$. Its in-plane matrix $D_d=\operatorname{diag}(-1,1)$
imposes $\bm\alpha^s=-D_d\bm\alpha^sD_d^T$ and permits only off-diagonal spin
responses. The $E_u(Q_s)$ structure retains the spin-preserving twofold rotation,
whose matrix $D_s=\operatorname{diag}(1,-1)$ imposes
$\bm\alpha^s=D_s\bm\alpha^sD_s^T$ and permits only diagonal spin responses.
Thus
\begin{subequations}
\label{eq:spin-thermoelectric-tensors}
\begin{align}
\bm\alpha_{Q_d}^{s}
&=\begin{pmatrix}0&\alpha_{xy}^{s}\\ \alpha_{yx}^{s}&0\end{pmatrix},
\label{eq:spin-tensor-qd}\\
\bm\alpha_{Q_s}^{s}
&=\begin{pmatrix}\alpha_{xx}^{s}&0\\ 0&\alpha_{yy}^{s}\end{pmatrix}.
\label{eq:spin-tensor-qs}
\end{align}
\end{subequations}
The charge tensors carry no spin-exchange minus sign and are diagonal in both
states. For $-\bm\nabla T\parallel x$, $Q_d$ therefore selects a transverse
pure spin current, while $Q_s$ selects a longitudinal spin-polarized current
[Figs.~\ref{fig:transport}(a) and \ref{fig:transport}(b)].

The calculated tensors follow the component pattern allowed above. For $E_u(Q_d)$,
$\alpha_{xx}^{\uparrow}=\alpha_{xx}^{\downarrow}$, so the longitudinal current
is unpolarized [Fig.~\ref{fig:transport}(c)]. In contrast,
$\alpha_{yx}^{\uparrow}=-\alpha_{yx}^{\downarrow}\neq0$ makes the transverse
charge contributions cancel and the spin contributions add, producing a pure
spin current [Fig.~\ref{fig:transport}(e)]. For $E_u(Q_s)$,
$\alpha_{xx}^{\uparrow}\neq\alpha_{xx}^{\downarrow}$ produces a longitudinal
spin-polarized current [Fig.~\ref{fig:transport}(d)], while
$\alpha_{yx}=0$ removes the transverse response [Fig.~\ref{fig:transport}(f)].
Figures~\ref{fig:transport}(g) and \ref{fig:transport}(h) rotate the $E_u$
order parameter at fixed $|\bm Q|=(Q_d^2+Q_s^2)^{1/2}$. Since
$Q_s\propto E_x$ and $Q_d\propto E_y$, this scan keeps the in-plane pump-field
magnitude fixed and varies only its direction. The transverse
$\alpha_{yx}^{s}$ requires an $E_y$ component and vanishes for pure $E_x$
driving [Fig.~\ref{fig:transport}(g)]. The longitudinal $\alpha_{xx}^{s}$
instead requires an $E_x$ component and vanishes for pure $E_y$ driving
[Fig.~\ref{fig:transport}(h)], matching the $Q_d$ and $Q_s$ tensor forms above.
Thus, at fixed phonon frequency and fixed $|\bm Q|$, rotating only the pump
polarization continuously tunes the spin current between transverse and
  longitudinal spin-polarized currents.

The $[U\parallel\mathcal I]$ operation of the $Q=0$ structure relates the complete
spin-resolved band structures at $\bm Q$ and $-\bm Q$ while exchanging
the spin channels. The spin-thermoelectric response therefore obeys
$\bm\alpha^{s}(-\bm Q)=-\bm\alpha^{s}(\bm Q)$. For a fixed temperature
gradient, $\bm J^s$ has the same odd parity. Reversing
the phonon coordinate therefore reverses both the splitting and the associated
spin current. Applying the frozen-mode coefficients to the trajectories in
Fig.~\ref{fig:dynamics} gives the time-dependent thermoelectric coefficients
$\bm\alpha^s[\bm Q(t)]$ reported in Fig.~S4 of the Supplemental
Material~\cite{SupplementalMaterial}. Each coefficient gives the corresponding
charge-current-equivalent spin current per unit temperature gradient. For a
specified gradient, the current is
$J_i^s(t)=\alpha_{ij}^s[\bm Q(t)](-\nabla_jT)$.
Because $\bm\alpha^s(-\bm Q)=-\bm\alpha^s(\bm Q)$, a coherent phonon produces
an alternating spin current with the phonon period.

\textit{General symmetry classification.}---To generalize phonon-controlled
symmetry selection beyond monolayer MnPSe$_3$, we determine when a single
$\Gamma$-point distortion can produce spin splitting by enumerating all 448
two-dimensional collinear spin layer groups without spin--orbit coupling and
at fixed collinear order~\cite{Du2026SpinLayerGroups}. Among them, 276 enforce
spin degeneracy throughout the Brillouin zone. For these 276 groups, we
first note that $\Gamma$-point phonons in a two-dimensional layer group are
either nondegenerate modes or doubly degenerate phonon doublets. We enumerate
each nondegenerate coordinate and, for each doublet, the symmetry-inequivalent
directions in its two-component coordinate space, obtaining 1300 combinations.
Among them, 384 remove full-Brillouin-zone spin degeneracy and permit splitting
at generic momentum. Of these 384 cases, 194 involve
infrared-active irreducible representations of the undistorted structure, 78
involve representations that are Raman active but not infrared active, and 112
involve representations silent in both linear optical responses. The full
classification is given in the Supplemental
Material~\cite{SupplementalMaterial}.
This classification links each active path to the corresponding $Q=0$ spin layer group and a
$\Gamma$-point phonon coordinate, and specifies the resulting subgroup, nodal
class, optical activity, and required pump-polarization direction.

Rectangular monolayer $r$-CrF$_3$ illustrates a different symmetry origin of
spin degeneracy~\cite{Chen2025CrF3}. Although its $Q=0$ antiferromagnetic state
lacks $\mathcal{PT}$ symmetry, the opposite-spin sublattices are connected by
$[U\parallel\{C_{2z}|\bm\tau\}]$ and
$[U\parallel\{M_z|\bm\tau\}]$. These two spin-exchanging operations contain
the fractional translation $\bm\tau$ and enforce spin degeneracy throughout
the Brillouin zone. Applying the same two mode-selection conditions used for
MnPSe$_3$, we find that the Raman-active $B_{2g}$ and $B_{3g}$ distortions
remove both operations. First-principles bands show $s$-wave splitting for
$B_{2g}$ and $d$-wave splitting for $B_{3g}$ [Fig.~S5 in the Supplemental
Material~\cite{SupplementalMaterial}]. The $r$-CrF$_3$ calculation therefore
shows that phonon-induced spin splitting also occurs when the $Q=0$ degeneracy
is enforced by nonsymmorphic spin-exchanging operations and the selected modes
are Raman active. Their residual groups further select complementary
spin-thermoelectric tensor components: $B_{2g}$ permits diagonal spin responses,
whereas $B_{3g}$ permits off-diagonal spin responses while forbidding a
transverse charge response. Polarization-selective Raman excitation would
therefore enable symmetry-controlled programming of $s/d$-wave splitting together
with the symmetry-allowed longitudinal/transverse spin-current responses.
Symmetry-compatible examples include the isostructural compound MnPS$_3$
and other two-dimensional antiferromagnets such as Hf$_2$S and
MgCr$_2$O$_4$~\cite{Chu2020MnPS3,Bai2025Hf2S,Tian2026MgCr2O4}.

\textit{Conclusion.}---Coherent $\Gamma$-point phonons activate hidden
unconventional magnetism in spin-degenerate collinear antiferromagnets by
breaking selected spin-layer operations without changing the N\'eel order. In
monolayer MnPSe$_3$, the pump frequency selects $A_{2u}$ or $E_u$, while the
in-plane polarization within the $E_u$ resonance selects $Q_d$ or $Q_s$. The
resulting $i$-, $d$-, and $s$-wave splittings are odd in $Q$, and the two
$E_u$ directions tune the spin-thermoelectric response between transverse
pure-spin and longitudinal spin-polarized currents. The complete cSLG
classification, together with the $r$-CrF$_3$ example, shows that
phonon-controlled symmetry selection extends beyond $\mathcal{PT}$-protected
systems and infrared-active modes to nonsymmorphic systems and Raman-active modes, establishing a symmetry-based
rule for programming spin splitting and its transport readout in two
dimensions.

\section*{Author Contributions}
X.S. performed the calculations, analyzed the results, and wrote the first
draft. C.T. and W.R. discussed the results and revised the manuscript. W.R.
designed and supervised the project.

\section*{Acknowledgments}
W.R. acknowledges funding from the bilateral CNR/NSFC project CINA 2024--2025,
``Ferroelectric and chiral hybrid perovskites'' (CUP B53C23007160005), and
support from the National Natural Science Foundation of China (Grant Nos.
12311530675 and 52130204). C.T. acknowledges support from the Science and
Technology Commission of Shanghai Municipality (Grant Nos. 25ZR1402162 and
25511103400).

\section*{Data Availability}
The datasets generated and analyzed in this study are not publicly available
because of ongoing related research and associated intellectual-property
considerations, but are available from the corresponding author upon
reasonable request.

\nocite{Kresse1996VASP,Kresse1999PAW,Perdew1996PBE,Dudarev1998DFTU,
Togo2015Phonopy,Togo2023Phono3py,Pizzi2020Wannier90,Pizzi2014BoltzWann}
\bibliography{references}

\end{document}